\documentclass{article}[10pt]

\usepackage{amsfonts}
\usepackage{amsmath}
\usepackage{amssymb}

\begin{document}
                                                                
\title{Analytical solutions for two inhomogeneous cosmological models with energy flow and dynamical curvature}

\author{Peter C. Stichel\\
Fakult\"at f\"ur Physik,  Universit\"at Bielefeld\\
D-33501 Bielefeld, Germany\\
email:  peter@physik.uni-bielefeld.de}
%\date{22.05.2018}

\maketitle

\begin{abstract}
Recently we have introduced a nonrelativistic cosmological model (NRCM) exhibiting a dynamical spatial curvature. For this model the present day cosmic acceleration is not attributed to a negative pressure (dark energy) but it is driven by a nontrivial energy flow leading to a negative spatial curvature. In this paper we generalize the NRCM in two different ways to the relativistic regime and present analytical solutions of the corresponding Einstein equations. 
These relativistic models are characterized by two inequivalent extensions of the FLWR metric with a time-dependent curvature function 
$K (t)$ and an expansion scalar $a(t)$. The fluid flow is supposed to be geodesic. The model V1 is shear-free with isotropic pressure and therefore conformal flat. It shows some common properties with the spherically symmetric Stephani models but it exhibits also some specific differences. In contrast to V1 the second model V2 shows a nontrivial shear and an anisotropic pressure.
For both models the inhomogeneous solutions of the corresponding Einstein equations will agree in leading order at small distances with the NRCM if a(t) and K(t) are each identical with those determined in the NCRM. This will be achieved by the demand of vanishing isotropic pressure and its first derivative w.r.t. $r^2$ at the coordinate origin $r = 0$.
Then the metric is completely fixed by three constants. The arising energy momentum tensor contains a nontrivial energy flow vector. 
Our models violate locally the weak energy condition.  As this may be caused by some averaging we speculate about to view each of our models as a local average of some other more fundamental model.  Global volume averaging leads to explicit expressions for the effective scale factor and the expansion rate $H (z)$.  Backreaction effects cancel each other for the model V2 but they are nonzero and proportional to the square of the magnitude of the energy flow for the model V1.
 The large scale (relativistic) corrections to the NCRM results are small for the model V2 for a small-sized energy flow. We have reproduced a corresponding adjustment of the three free constants from [1] to cosmic chronometer data leading to the prediction of an almost constant, negative value for the dimensionless curvature function $k(z) \sim - 1$ for redshifts $z < 2$.   
\end{abstract}

\section{Introduction}

There is no doubt that the present Universe goes through a phase of (real or apparent) accelerated expansion (see [2] and the literature cited therein). According to recent claims almost all observations are in good agreement with the standard $\Lambda$CDM-cosmological model, only some  disagreements are recorded (see [3]).  But in some very recent papers [4], [5] even the validity of the standard model has been put into question. 

Two alternative strategies to the standard model are under discussion.

In the first category one introduces some kind of ''new physics'' by changing Einstein's equations (EEs) either by modifying the geometrical part of the EEs (called modified gravity), or by changing the matter part.

In the second category one considers the accelerated expansion as an apparent effect due to averaging over inhomogeneities in the Universe (called backreaction, see [6] for a recent review).

For cosmological models based on averaging over inhomogeneities, one comes to the conclusion that the present day cosmic acceleration is due to a negative spatial curvature [7], [8].  A comparison of such backreaction effects with observations has been undertaken in [9]. Furthermore numerical solutions of Einstein's equations for a Silent Universe show ''that the spatial curvature emerges due to nonlinear evolution of cosmic structures'' [10].

Inhomogeneous cosmological models containing a time-dependent curvature function have been first introduced by Stephani [11] (see also [12] and the literature cited therein).  Stephani models are the most general conformal flat perfect fluid solutions of Einstein's equations with nontrivial expansion [12]. They are characterized by an inhomogeneous pressure and a homogeneous energy density. If one chooses a Friedman-like time coordinate [13] the model contains two unknown functions of time, an expansion scalar $a(t)$ and a curvature function $K(t)$. For a subclass of the spherical symmetric case $K(t)$ and $a(t)$ are postulated to be proportional to each other (called Dabrowski models [14]). But this assumption fixes the sign of the curvature function $K(t)$. Quite recently comparisons of such models with observational data have been performed in [15] and [16]. It turns out that such models are only observationally acceptable if the inhomogeneity parameter is assumed to be small and the model contains in addition a standard dark energy component [15], [16].

Recently we have shown that a nonrelativistic cosmological model (NRCM) introduced in [17], reviewed in [18] and derived as the nonrelativistic limit (approximation at sub-Hubble scales) of a general relativistic model exhibits a dynamical curvature function with a negative value at the present cosmological epoch [19], [20].  In a very recent paper [1] we have fixed the three constants (initial conditions) of the model by adjusting them in two different ways to a second order polynomial fit by Montanari and R\"as\"anen [21] to the observed expansion rate $H(z)$. In the particular case of scenario 2 in [1] we obtain for the dimensionless curvature function the prediction $k(z) \sim - 1$ for  $z < 2$.

In the present paper we consider two different relativistic generalizations of the NRCM. These models are characterized by two inequivalent extensions of the FLWR metric with a time dependent curvature function and a geodesic fluid flow. Each of both models contain again two functions of time, a scale factor $a(t)$ and the curvature function $K(t)$. The solutions of the corresponding Einstein equations turn out to be inhomogeneous. They will agree in leading order at small distances with the NRCM if $a(t)$ and $K(t)$ are each identical with those found in the NRCM. Therefore the metric as well as the energy momentum tensor become completely fixed. 

It turns out that our models violate locally the weak energy condition.  But due to the fact that averaging may lead to a violation of energy conditions [6] we speculate about to view each of our models as a local average of some other yet unknown but more fundamental cosmological model.

To compare the outcome of our models with observational results we consider a global volume average of our inhomogeneous analytical solutions.

The paper is organized as follows: In section 2 we summarize the essentials of the NRCM and recapitulate the determination of the three free constants from scenario 2 in [1]. In section 3 we describe the two relativistic generalizations. Variant 1 consists of a shear-free fluid with isotropic pressure (subsection 3.1). This model will be compared with the spherically symmetric Stephani models in subsection 3.1.1.
A shearing model with anisotropic pressure (variant 2) will be discussed in subsection 3.2. Spatial averaging, backreaction and the size of the large-scale (relativistic) corrections will be considered in section 4. We finish the paper with some concluding remarks (section 5).

\section{Summary of the nonrelativistic cosmological model (NRCM)}

In the following we summarize the essentials of our NRCM with dynamical curvature. We will present only the resulting equations but give a full account of their physical interpretation.

Here and throughout the whole paper we will consider only spherical symmetric geometry. Then the fluid flow is irrotational.

We start with Einstein's equations (EEs) for a self-gravitating  geodesic fluid (velocity field $u^\mu$; we use units $c = 1 = 8 \pi G$)  
\begin{equation}
G_{\mu\nu} = T_{\mu\nu}
\end{equation}
with an energy-momentum tensor (EMT) containing in the comoving frame only energy density and an energy  flow vector $q_\mu (u^\mu q_\mu = 0)$
\begin{equation}
T_{\mu\nu} = \rho u_\mu u_\nu + q_\mu u_\nu + q_\nu u_\mu \ .
\end{equation}
In  the nonrelativistic  and shear-free limit [19] (or at small distances [20]) we obtain from the EEs, after having eliminated the energy flow vector, the following system of two coupled ordinary differential equations for the cosmological scale factor $a(t)$ and the active gravitational mass density (for any details we refer to [19] and [20] respectively).                   
\begin{equation}
\rho = - \frac{6\ddot{a}}{a}
\end{equation}                
and
\begin{equation}
\dot{\rho} + 3 \frac{\dot{a}}{a} \rho + \frac{6 K_1}{a^5} = 0
\end{equation}
where the constant $K_1$ measures the strength of the energy flow (see [1]).
                                                                  
For the curvature function $K(t):= \frac{a^2}{6} R^\ast$  ($R^\ast$  is the spatial curvature) we obtain from the Hamiltonian constraint (we define $\hat{\rho} : = \frac{a^3 \rho}{6}$)
\begin{equation}
K(t) = - \dot{a}^2 + \frac{2\hat{\rho}}{a}\ .
\end{equation}
In the limit of vanishing energy flow $(K_1  = 0)$ our model reduces to the flat FLRW dust model.
The dynamical system (3, 4) possesses two constants of motion $Q_i$ ($i = 2, 3$)        
\begin{equation}
Q_2 = K_1 \dot{a} - \frac{1}{2} \hat{\rho}^2, ~~~Q_3 = - \frac{\hat{\rho}^3}{6} - Q_2 \hat{\rho} + \frac{K_1^2}{a}\ .
\end{equation}
On the solution space of (3, 4) the $Q_i$   take constant values   $K_i$   which are determined by the initial value of $\rho$ and by the Hubble parameter.
                            .  
Introducing the redshift $z = a^{-1} - 1$ (the validity of this relation will be discussed in subsection 4.1) instead of the time $t$ as independent variable, we get finally from (6) analytic expressions for the expansion rate $H(z) : = \dot{a}/a$ and for the curvature function $K(z)$. In dimensionless units
\begin{eqnarray}
k_1 : = \frac{K_1}{H^3_0} &,& k_2 : = \frac{K_2}{H^4_0}~~ , ~~~k_3 : = \frac{K_3}{H^6_0}\nonumber\\
& & h(z) : = \frac{H(z)}{H_0} ,~~~ k(z) : = \frac{K(z)}{H^2_0}
\end{eqnarray}
we get [1]

\begin{description}
\item{$\bullet$} A cubic equation for $h(z)$
\begin{equation}
(k^2_1 (1+z) - k_3)^2 = \frac{2}{9} \left( \frac{k_1 h(z)}{1+z} - k_2 \right) \left( \frac{k_1 h(z)}{1+z} + 2k_2\right)^2
\end{equation}                                                                                                
 
\item{$\bullet$} and $k(z)$ in terms of $h(z)$ 
\begin{equation}
k(z) = - \left( \frac{h(z)}{1+z}\right)^2 \pm 2^{3/2} (1+z) \left( \frac{k_1 h(z)}{1+z} - k_2 \right)^{\frac{1}{2}}
\end{equation}
with the  $+$  sign for $z > z_t$ and the -- sign for $z <  z_t$
\end{description}

\noindent 
where $z_t$  defines the transition redshift  (given in terms  of the $k_i$ by (12)).
 
The constants $k_i$ are related to observable quantities as follows:
\begin{description}
\item{$\bullet$} $k_1$ determines the magnitude of the derivative of the curvature function $k(z)$ with respect to $z$ 
\begin{equation}
k^\prime (z) = \frac{2k_1 (1 + z)^2}{h(z)}
\end{equation}
\item{$\bullet$} the deceleration parameter $q_0 : = q(0) (q(z) : = - \frac{\ddot{a}}{a H^2})$ is given in terms of $k_1$ and $k_2$ by
\begin{equation}
q_0 = - \sqrt{2} (k_1 - k_2)^{1/2}
\end{equation}
\item{$\bullet$} $z_t + 1$ is proportional to $k_3$
\begin{equation}
z_t + 1 = k_3/k^2_1 \ .
\end{equation}
\end{description}

The $k_i$ are not independent of each other. By taking (8) at $z = 0$ we obtain the relation
\begin{equation}
(k_3 - k^2_1)^2 = \frac{2}{9} (k_1 - k_2) (k_1 + 2k_2)^2
\end{equation}
leading to the constraint
\begin{equation}
k_1 > k_2\ .
\end{equation} 
What about the signs of the $k_i$?  We require that for $z > z_t$   our model describes a decelerating phase and for $z < z_t$ an accelerating phase of the Universe.  Then, according to (3) and (6), $k_2$  and $k_3$ must be positive [17], [18] leading by the constraint (14) to $k_1 > 0$.

We conclude [19], [20]:
\begin{description}
\item{$\bullet$} According to our model the present day cosmic acceleration is not attributed to any kind of negative pressure (dark energy) but it is driven by a nontrivial, radially directed energy flow.  
\item{$\bullet$} At least for $z < z_t$   we have, according to (9), a negative spatial curvature (hyperbolic space).  But $k(z)$ is an increasing function of $z$ and we have $k(z_t) < 0$. Therefore at some
 $z = z_0 > z_t$ we should have a transition from a hyperbolic to a spherical space.
\item{$\bullet$}	According to (3) the energy density turns out to be positive in the decelerating phase and negative in the accelerating phase of the Universe.  So at present the weak energy condition is violated in our nonrelativistic model [20]. But the situation becomes more complicated for the relativistic generalizations of our model, which are of an inhomogeneous type (see section 3). Hence we postpone the discussion of the validity/violation of energy conditions to section 3.
\end{description}

\subsection{Estimation of the $k_i$ by observations}
 
We have to fix two independent constants $k_i$ and the Hubble parameter $H_0 := H(0)$   by means of some data for the expansion rate $H(z)$. The only data for $H(z)$ which are independent of any cosmological model are the cosmic chronometer data (see table 1 in [21]). But these data possess still rather large error bars. Hence we dispense with a least-squares fir. Instead we try to get a reasonable fit with a very small value of $k_1$ which would keep the relativistic corrections for the model V2 small (see section 4). This has been done for scenario 2 in (1) by using the value of $H(z)$ at decoupling leading to 
\begin{equation}
k_1  = 0.002082 
\end{equation}
Here we adopt this value as a viable estimate for $k_1$ and follow for the determination of $k_{2,3}$ the procedure in [1]:

From (8), taken at $z = z_t$, we obtain
\begin{equation}
\frac{k_2}{k_1} = \frac{h(z_t)}{1 + z_t} \ .
\end{equation}

By inserting the second order polynomial fit [22] to the cosmic chronometer data (table 1 in [21])
\begin{equation}
 h(z) = h_1 z + h_2  z^2 ~~~~~~ \mbox{with}~~ h_1  = 0.8368 ~~~~~~  \mbox{and}~~ h_2 = 0.1082 
 \end{equation}
into (16) and taking (13) together with $k_1 $  from (15) into account we obtain a coupled system of two algebraic equations for $z_t$ and $k_2$. From its solution and the relation (12) we obtain finally
\begin{equation}
k_2  = 0.0020773 ~~~~~~~   \mbox{and} ~~~~~~~  k_3 = 1.072 \times 10^{-5} \ .
\end{equation}
With these values for the $k_i$   we get from (8) numerical results for $h(z)$ which are compared with observational data and the polynomial fit (17) in table 1.

In addition we have listed our predictions for the curvature function $k(z)$ which show an almost constant behavior $k(z) \sim - 1$ for all $z < 2$ in agreement with the FLRW consistency conditions [21].

\bigskip
\noindent
\begin{tabular}{l | r | r |r | c | r }
\hline                                                                                 
  z     &      $H_{ob}(z)$ &  $\sigma_H$       &    $H_{pol}(z)$   &                     $H(z)$     &           $k(z)$\\ \hline

0.07    &      69   & 19.6          &       67.995    &                 68.679  &      - 1.00582\\
0.12    &      68.6 & 26.2        &       70.747    &                        71.880  &      - 1.00559\\
0.179   &      75    & 4        &       74.039    &                    75.653  &     - 1.00531\\
0.199   &      75   & 5         &       75.166    &                               76.932  &     - 1.00521\\
0.2     &      72.9 & 29.6     &       75.222    &                              76.996  &    - 1.00520\\
0.28    &      88.8 & 36.6      &       79.787    &                              82.112  &    - 1.00478\\
0.352   &      83   & 14~~        &       83.971    &                        86.715  &    - 1.00438\\
0.3802  &      83   & 13.5        &       85.629    &                           88.518  &    - 1.00422\\
0.4004  &      77    & 10.2       &       86.824    &                        89.810  &    - 1.00410\\
0.4247  &      87.1 & 11.2        &       88.269    &                           91.363  &    - 1.00396\\
0.4497  &      92.8  & 12.9        &       89.764    &                          92.960  &    - 1.00381\\
0.4783  &      80.9  & 9.0         &       91.485    &                    94.789  &    - 1.00363\\
0.48    &      97    & 62      &       91.587    &                            94.898  &    - 1.00362\\
0.593   &      104  & 13         &       98.500    &                           102.121 &     - 1.00289\\
0.680   &      92    & 8        &       103.943   &                           107.682 &     - 1.00228\\
0.781   &      105  & 12        &       110.394   &                       114.130 &     - 1.00154\\
0.875   &      125  & 17        &       116.526   &                       120.147 &     - 1.00081\\
0.88    &      90   & 40      &       116.855   &                            120.466 &     - 1.00077\\
1.037   &      154  & 20        &       127.380   &               130.505 &     - 0.99945\\
1.363   &      160  & 33.6      &       150.329   &               151.362 &     - 0.99559\\
1.965   &      186.5 & 50.4       &       196.587   &               189.969 &     - 0.99004\\
\hline\hline
\end{tabular}

\medskip
\noindent
{\bf Table 1:} Expansion rate data $H_{ob}$ (2nd column, taken from table 1 in [21] with errors $\sigma_H$) versus polynomial fit $H_{pol}$ [21](4th column; see eq. (17)) and predictions for $H(z)$ (5th column). For the polynomial fit as well as for the predictions we have used $H_0 = 64.2 km/s/Mpc$  [21].  The 6th column shows the predictions for the curvature function $k(z)$.

\bigskip
\noindent
Our results show that the observation of such an almost constant behavior of $k(z)$ is not automatically an evidence for the validity of the FLRW model where a constant $k$ is an ad hoc parameter. Of course, the FLRW model is one possible model showing such a behavior, but it is not the only one.  

In contrast to the cosmic chronometer data for $H(z)$ the use of SN  Ia-data and BAO-data respectively rests upon a distance-redshift relation which depends heavily on the underlying cosmological model . For a generic model the area distance satisfies a differential equation (see eq. (49) in [23]). In [19] we have adjusted this equation to our nonrelativistic model but we were not able to solve it analytically. To proceed, one has to take the relativistic (inhomogeneous) corrections (see section 3) into account and uses numerical methods (or at least an expansion in powers of z for low redshifts; see [24] for the case of the Stephani models). But this is outside the scope of the present paper. 

To compare our model with the measurement of the CMB-shift parameter we should have some knowledge on the matter part of the energy density. But our model does not allow for the separation of the energy density into a matter and a dark energy part [17].

\section{Relativistic generalizations of the NRCM}

Unfortunately we did not succeed to get an analytic solution for the full relativistic EES with the EMT (2) in spherical symmetric space-time. But in order to get in leading order at small distances the results described in section 2, it is not necessary to use the EMT (2).  Instead we may start with an extended EMT containing in addition nontrivial pressure terms                                                            
\begin{equation}
T_{\mu\nu} = (\rho + p_t) u_\mu u_\nu + p_t g_{\mu\nu} + (p_r - p_t) s_\mu s_\nu + q_\mu u_\nu + q_\nu u_\mu
\end{equation}                   
where $p_r$ denotes the radial pressure,  $p_t$    the transversal pressure and $s_\mu$  is an unit space-like vector with $u^\mu s_\mu = 0$.
                       .                                    
But then the pressure terms must behave for small distances such, that in leading order the NRCM results (see section 2) are valid. Of course such a requirement has no unique answer. To realize it we have to distinguish between two options:

\begin{description}
\item{$\bullet$} Variant 1 (V1): The fluid flow is shear-free.

Our aim is to find a solution of the EEs with a time varying curvature function.  But, as shown in [19], this requires for a shear-free and geodesic fluid flow necessarily nontrivial pressure terms in the EMT.

In the following section 3.1 we consider only the case of isotropic pressure. 

\item{$\bullet$} Variant 2 (V2):  The fluid flow has non-vanishing shear.

This case will be considered in section 3.2.
\end{description}

\subsection{Shear-free model with isotropic pressure (V1)}

We consider only the case of isotropic pressure $(p_r = p_t =: p)$. Then the EMT (19) may be written as
\begin{equation}
T_{\mu\nu} = \rho u_\mu u_\nu + p h_{\mu\nu} + q_\mu u_\nu + q_\nu u_\mu
\end{equation}
where  $h_{\mu\nu} : = u_\mu u_\nu + g_{\mu\nu}$  projects onto the space orthogonal to $u^\mu$.

The geodesic and shear-free fluid motion allows the consideration of a co-moving coordinate system with $u^\mu = \delta^\mu_0$  which may be written as (see [25])
\begin{equation}
d s^2 = - d t^2 + V^{-2} (t,r) (d r^2 + r^2 d \Omega^2) \ .
\end{equation}
The condition of the isotropy of the pressure, inserted into the EEs (1), leads to a differential equation for $V$ with respect to $x: = r^2$ [26], [27]
\begin{equation}
\frac{d^2}{d x^2} V (t,x) = 0 
\end{equation}
whose solution is conveniently be written as [27]
\begin{equation}
V^{-1} (t,r) = \frac{a(t)}{1 + \frac{r^2}{4} K(t)} \ .
\end{equation} 
We note that vanishing anisotropy of the pressure implies that the space-time is conformal flat (see [25]).

Insertion of (23) into the line element (21) leads to a metric which differs from the usual FLRW metric only by the time-dependence of K [27].

For the moment the functions $a(t)$ and $K(t)$ are  not yet specified, but later we will see that they can be identified with the corresponding functions of the NRCM (see section 2).

Introducing the metric (21) with (23) together with the EMT (20) into the EEs (1) we obtain (see [27])
for the energy flow $q~~(q_\mu = q s_\mu, s_\mu = (0,1/V))$
\begin{equation}
q = - \frac{r \dot{K}}{a (1 + \frac{r^2}{4} K)}
\end{equation}
for the energy density $\rho$
\begin{equation}
\rho = 3 \frac{K}{a^2} + \frac{1}{3} \Theta^2
\end{equation}
and for the pressure $p$
\begin{equation}
p = - \frac{2}{3} \dot{\Theta} - \frac{1}{3} \Theta^2 - \frac{K}{a^2}
\end{equation}
with the volume expansion $\Theta : = \nabla_\mu u^\mu = -3 \frac{\dot{V}}{V}$ given by
\begin{equation}
\Theta = 3 \left( \frac{\dot{a}}{a} - \frac{\frac{r^2}{4} \dot{K}}{1 + \frac{r^2}{4} K} \right)
\end{equation}
where a dot represents differentiation with respect to time $t$.

So, the spatial scalar curvature $R^\ast$, obtained from the Hamiltonian constraint $R^\ast = 2 \rho - \frac{2}{3} \Theta^2$,
turns out to be only a function of time (see [28], [25])
\begin{equation}
R^\ast = \frac{6 K(t)}{a^2 (t)} \ .
\end{equation}
Let us now look at the behavior of the dynamical quantities at small distances $r$.

For the energy density $\rho$ we obtain from (25) and (27)
\begin{equation}
\rho \mathop{=}_{r\to 0} 3 \left( \frac{K}{a^2} + \left( \frac{\dot{a}}{a} \right)^2 \right) + 0 (r^2)\ .
\end{equation}
To bring (21) in leading order for small distances into agreement with the NRCM result (3) we have to require
\begin{equation}
K = - (2 \ddot{a} a + \dot{a}^2) \ .
\end{equation}
Next we have to consider the local energy conservation equation
\begin{equation}
\dot{\rho} + \Theta (\rho + p) + \frac{1}{r^2 B^3} (r^2 q B^2)^\prime = 0
\end{equation}
where a prime represents differentiation with respect to $r$.

The last term in (23), the expansion $\Theta$ and the pressure $p$ respectively behave at small distances as
\begin{equation}
\frac{1}{r^2 B^3} (r^2 B^2)^\prime \mathop{=}_{r\to 0} - \frac{3\dot{K}}{a^2} + 0 (r^2),~~~~\Theta \mathop{=}_{r\to 0} 3 \frac{\dot{a}}{a} + 0 (r^2)
\end{equation}
and
\begin{equation}
p = - \left( 2 \frac{\ddot{a}}{a} + \left( \frac{\dot{a}}{a} \right)^2 + \frac{K}{a^2} \right) + \frac{r^2}{2} \left( 3 \frac{\dot{a}}{a} \dot{K} + \ddot{K} \right) + 0 (r^4) \ .
\end{equation}

Therefore, to bring (31) at leading order at small distances in agreement with the NRCM result (4) we have to require according to (32), (33) and taking into account (30)
\begin{equation}
\dot{K} = - \frac{2 K_1}{a^3} ~~~~, ~~~~~~~~~~~~K_1 = \mbox{const.} ~~.
\end{equation}
From the behavior of the pressure for small distances, eq. (33), we observe that the conditions (30) and (34) may be formulated exclusively in terms of $p(t, x)$ $(x : = r^2)$ for $x \to 0$          
\begin{equation}
p (t,0) = 0 ~~~~~~~~~~~~~~~\mbox{and} ~~~~~~~~~~~~~ p^\prime (t,0) = 0
\end{equation}
Combining (30) and (34) we obtain
\begin{equation}
(\ddot{a} a^2)^\cdot = \frac{K_1}{a^2}
\end{equation}  
which is identical with the differential equation for $a(t)$ obtained in the NRCM (eliminate $\rho$ in (4) by means of (3)).

The aforementioned results lead finally to the following {\bf conclusions}: 

\begin{description}
\item{$\bullet$} Variant 1 of the relativistic model agrees in leading order at small distances with the NRCM if and 
only if the pressure $p(t, x)$ satisfies the conditions (A)
$$
p(t, 0) = 0 ~~~~~~~~~(A1)~~~~~~~~~~~~~~~~~~~~and~~~~~~~~~~~~~~~~~~p^\prime (t, 0) = 0~~~~~~~~~~~(A2)
$$
which implies (36). So, due to the results presented in section 2 (see equations (8) and (9)) we obtain exact analytic expressions for  $\dot{a}$ and $K$ as functions of the NRCM scale factor $a$. Then $a(t)$ may be obtained by quadrature (see [17], appendix A).
\end{description}
 
\bigskip
\noindent
{\bf Conclusion:}

With the known expressions for $a(t)$ and $K(t)$ our metric (21), (23) is completely fixed in terms of the three constants $K_i (i = 1,2,3)$. Therefore, 
by using (30) and (34) in (24) -- (26) we obtain the following inhomogeneous solution of Einstein's equations
\begin{equation}
q = \frac{4}{ar} f
\end{equation}
\begin{equation}
\rho = 3 \left( \frac{K}{a^2} + \left( \frac{\dot{a}}{a} + f \right)^2 \right)
\end{equation}
and                                                                                         
\begin{equation}
p = - 5 f^2
\end{equation}
where we have defined
\begin{equation}
f (t,r) : = \frac{K_1 \frac{r^2}{2}}{a^3 (t) (1 + \frac{r^2}{4} K(t))}
\end{equation}
In terms of  $f$ we get for the volume expansion 
\begin{equation}
\Theta = 3 \left( \frac{\dot{a}}{a} + f\right) \ .
\end{equation}

Readers who are more familiar with the system of evolution and constraint equations (see [25] for our case) instead of the EEs may easily check that the results (37) -- (39) and (41) for $q, \rho, p$  and $\Theta$  together with (30) and (34) identically satisfy the conservation equations
\begin{equation}
\dot{\rho} + \Theta (\rho + p) + \frac{q^\prime}{B} + 2 q \frac{(r B)^\prime}{r B^2} = 0
\end{equation}
and 
\begin{equation}
\dot{q} + \frac{p^\prime}{B} + \frac{4}{3} \Theta q = 0
\end{equation}
the Raychaudhuri-Ehlers equation
\begin{equation}
\dot{\Theta} + \frac{1}{3} \Theta^2 + \frac{1}{2} (\rho + 3 p) = 0
\end{equation}
as well as the constraint equations                                                
\begin{equation}
\rho^\prime = \Theta q B
\end{equation}
and
\begin{equation}
\frac{2}{3} \Theta^\prime = q B
\end{equation}

Our results exhibit the following interesting features:
\begin{description}
\item{$\bullet$}	The large-scale (relativistic) corrections are determined by only one function $f (t, r)$ which is proportional to the energy flow, vanishes at the coordinate origin and becomes singular for $r \to 2 |K|^{-1/2} = : r_\infty (t)$ in case of a negative spatial curvature.
\item{$\bullet$}    The pressure turns out to be negative.
\item{$\bullet$}    The spatial curvature shows no relativistic correction (see  eq.(28)).
\end{description}
What about the sign of the energy density  $\rho$? Let us rewrite (38) by means of (30) as
\begin{equation}
\rho (t,r) = - \frac{6\ddot{a}}{a} + 3 f (2 \frac{\dot{a}}{a} + f)
\end{equation}

Then, in the accelerating phase of the Universe, $\rho$ turns out to be negative for small $r$ but it becomes singular, together with $f$, for $r\to r_\infty (t)$. Hence, by continuity, it exists for each fixed $t$ in the accelerating regime some finite value $r_0 (t) < r_\infty (t)$ such that 
\begin{equation}
\rho (t,r) < 0 ~~~ \mbox{for}~~ 0 \le r < r_0 (t) ~~~ \mbox{and}~~ \rho (t,r) > 0 ~~~ \mbox{for}~~~ r_0 (t) < r < r_\infty \ .
\end{equation}

 After (48) three remarks are in order:
\begin{description}
\item{$\bullet$} The weak energy condition (WEC) is locally (for small distances) violated.  This should not bother us as there are many viable general relativistic models violating some energy condition (e.g. scalar field coupling to gravity [29].

On the other hand it is well known that averaging may lead to a violation of energy conditions [6]. So one may speculate about to view our model as an average of another yet unknown model (which satisfies the WEC) over scales below the homogeneity scale of the Universe (local average [43]). But it is outside the scope of the present paper to discuss this idea in more detail.

\item{$\bullet$}	The division of space into one part containing a positive energy density and another part containing a negative energy density reminds us on the Dirac-Milne Universe [30], [31]. But it is outside the scope of the present paper to amplify this point.

\item{$\bullet$} Let us consider the subset of $r_0 < r < r_\infty$ for which the f-term in (38) dominates all other terms.  For r values in this subset we have
\begin{equation}
\rho \sim - 3/5 p ~~~~\mbox{and~therefore} ~~~\rho + 3p \sim \frac{12}{5} p < 0\ .
\end{equation} 
Then, according to the Raychaudhuri-Ehlers eq. (44), the accelerated expansion of the Universe is locally (for this subset) given by dark energy i.e. a negative pressure.  But the pressure p is proportional to the square of the magnitude of the energy flow.  So the primary cause for the accelerated expansion is still the nontrivial energy flow.
\end{description}

\subsubsection{Comparison with the spherically symmetric Stephani models}

The spherically symmetric Stephani models (SSSM) are defined by the metric [11] (see also [12] and [13])
\begin{equation}
d s^2 = - \frac{a^2}{\dot{a}^2} \left[ \frac{\dot{V}}{V}\right]^2 dt^2 + V^{-2} (dr^2 + r^2 d \Omega^2)
\end{equation}
where $V(t, r)$ is defined by (23) and we have chosen the Friedman-like time coordinate [13].

\medskip
\noindent
Our model V1 and the SSSM possess the following common properties:

\begin{description}
\item{$\bullet$} They are fully characterized by two time-dependent functions, an expansion scalar $a(t)$ and a curvature function $k(t)$.

\item{$\bullet$} They are shear-free, have isotropic pressure and are therefore conformal flat.

\item{$\bullet$} They are inhomogeneous and reduce for vanishing inhomogeneity to the flat FLRW dust model

\item{$\bullet$} In case of our model and for a subclass of the SSSM the pressure vanishes at the origin [13], [32].
\end{description}

\medskip
\noindent
The differences between both models are the following:

\begin{description}
\item{$\bullet$}	For the SSSM the energy density is homogeneous and the pressure is inhomogeneous. In case of our model the energy density and the pressure are both inhomogeneous.

\item{$\bullet$}	The fluid flow for our model is geodesic but it is non-geodesic in case of the SSSM.

\item{$\bullet$}	The energy flow is nontrivial for our model but it vanishes for the SSSM.

\item{$\bullet$}	The accelerated expansion of the Universe is driven primarily by the energy flow for our model but it is driven by the inhomogeneity in case of the SSSM [32], [33] [34].

\item{$\bullet$}	The expansion scalar a(t) and the curvature function k(t) are a priori free functions in case of the SSSM. For a subclass they are postulated to be proportional to each other. These are the so called Dabrowski models [14] (see also [15], [16], [13], [32], [33], [34])  which, however, do not allow a sign chance of k(t).  Contrary to the SSSM our model fixes both functions in terms of three constants $K_i$  by the NRCM dynamics (see section 2) which allows a dynamically determined sign change of k(t).
\end{description}                                                                                                                                                                                                                       
\subsection{Shearing model with anisotropic pressure (V2)}

We are not interested to study the most general shearing model with anisotropic pressure. Instead we will define our model by a metric which is as close as possible to the FLWR metric. This metric is given in co-moving coordinates with $u^\mu = \delta^\mu_0$ by
\begin{equation}
d s^2 = - d t^2 + a^2 (t) \left( \frac{dr^2}{1 - K(t) r^2} + r^2 d \Omega^2 \right) \ .
\end{equation}
This is the usual FLRW metric except that the constant curvature is replaced by a function of time $K(t)$. Insofar it is similar to the metric (21), (23) considered in section 3.1. Both are equivalent for K = const. as can be seen by the coordinate transformation 
\begin{equation}
r \to \frac{r}{1 + \frac{r^2}{4} K}
\end{equation}
which converts the metric (51) into the metric (21), (23). But for a time dependent curvature $K(t)$ both metrics are inequivalent [35]. The easiest way to see this is by considering the shear of the co-moving fluid, which for the metric (51) is different from zero and proportional to the time derivative of $K(t)$.

The metric (51) appeared already as a subcase of a more general metric in [35], later it has been considered as an effective metric for cosmological models with spatial averaging [36], [9].  

Introducing the metric (51) and the EMT (19) into the EEs (1) we obtain (see [37])
\begin{equation}
\rho = 3 \left( \left( \frac{\dot{a}}{a} \right)^2 + \frac{K(t)}{a^2} \right) + \frac{\dot{a}}{a} \frac{\dot{K} (t) r^2}{1 - K(t) r^2}
\end{equation}

\begin{equation}
p_r = - \frac{2\ddot{a}}{a} - \left( \frac{\dot{a}}{a} \right)^2 - \frac{K(t)}{a^2}
\end{equation}

\begin{eqnarray}
p_t &=& - 2 \frac{\ddot{a}}{a} - \left( \frac{\dot{a}}{a} \right)^2 - \frac{K (t)}{a^2} - \frac{3}{2} \frac{\dot{a}}{a} \frac{\ddot{K} (t) r^2}{1 - K(t) r^2}\nonumber \\
& & - \frac{3}{4} \frac{\dot{K} (t)^2 r^4}{(1- K(t) r^2)^2} - \frac{1}{2} \frac{\ddot{K}(t) r^2}{1 - K(t) r^2}
\end{eqnarray}
and
\begin{equation}
q = - \frac{1}{a} \frac{\dot{K} (t) r}{(1-K(t)r^2)^{1/2}}
\end{equation}

Furthermore, volume expansion $\Theta$ and shear $\sigma$, defined by the decomposition of the covariant derivative of $u_\mu ~~(\sigma^2 : = \frac{1}{2} \sigma_{\mu\nu} \sigma^{\mu\nu})$ 
\begin{equation}
\nabla_\mu u_\nu = \frac{1}{3} \Theta h_{\mu\nu} + \sigma_{\mu\nu}
\end{equation} 
are given by
\begin{equation}
\Theta = 3 \frac{\dot{a}}{a} + \frac{1}{2} \frac{\dot{K} r^2}{1 - K r^2}
\end{equation}
and
\begin{equation}
\sigma = \frac{1}{2 \sqrt{3}} \frac{\dot{K} r^2}{1 - K r^2} \ .
\end{equation}

Then the spatial curvature   $R^\ast$, obtained from the Hamiltonian constraint $R^\ast = 2 \rho - \frac{2}{3} \Theta^2 + 2 \sigma^2$, is again a function of time only and is given by the same expression (28) as for the model V1
\begin{equation}
R^\ast = \frac{6 K(t)}{a^2 (t)} \ .
\end{equation}
Now, by looking for the small distance behavior we could duplicate in detail the considerations from section 3.1 applied to the present case. We will not do so, but instead we shorten the discussion and start with the 

\bigskip
\noindent
{\bf Supposition:}

\medskip
\noindent  
Variant 2 of the relativistic model agrees in leading order at small distances with the NRCM if the total pressure 
$p (t, x)~~~~~ (p = 1/3 (p_r    + 2 p_t))$ satisfies the conditions (A)
$$
p (t, 0) = 0 ~~~~~~~~~~~ (A1)~~~~~~~~~ and ~~~~~~~~~ p^\prime (t, 0) = 0 ~~~~~~~~~~~~~~~~(A2)
$$

\newpage

\noindent
{\bf Proof:}

\smallskip
\begin{description}
\item{$\bullet$}	(A1) leads to (see (30))
\begin{equation}
K = - (2 \ddot{a} a + \dot{a}^2) \ ,
\end{equation}
and therefore the energy density  $\rho$ (53)   approaches for $r \to 0$ the expression given by eq. (3).

\item{$\bullet$}  The local energy conservation equation, which reads in our case (see eq. (34) in [38];
 $B^2 : = \frac{a^2}{1 - K r^2}$)
\begin{equation}
\dot{\rho} + (\rho + p_r) \frac{\dot{B}}{B} + 2 (\rho + p_t) \frac{\dot{a}}{a} + \frac{q^\prime}{B} + 2q \frac{1}{r B} = 0
\end{equation}
approaches for $r \to 0$, according to (A2) together with (61) and (56), the NRCM form eq. (4).  
\end{description}

Note that (A2) takes again the same explicit form as for the model V1 (see (33))
\begin{equation}
\ddot{K} + 3 \frac{\dot{a}}{a} \dot{K} = 0~~~~~~~~~\mbox{with~the~solution}~~~~~~\dot{K} = - \frac{2K_1}{a^3},~~~ K_1 = \mbox{const.}
\end{equation}
so that (61) together with (63) imply again the NRCM condition (36). Hence we get the same analytic expressions for a(t) and K(t) as for V1 and obtain finally the following  inhomogeneous solution of Einstein's equations
\begin{eqnarray}
\rho &=& - 6 \frac{\ddot{a}}{a} - 2 \frac{\dot{a}}{a} g\\
p_r &=& 0\\
p_t &=& - 3 g^2\\
\mbox{and}~~~~~~~~~~~~~~~~~~~~~~~ & & \nonumber\\          
q &=& \frac{2K_1 r}{a^4 (1 - Kr^2)^{1/2}}
\end{eqnarray}
where we have defined
\begin{equation}
g(t,r) : = \frac{K_1 r^2}{a^3 (t) (1-K(t) r^2)}\ .
\end{equation}
In terms of $g$ we obtain for the volume expansion
\begin{equation}
\Theta = 3 \frac{\dot{a}}{a} - g
\end{equation}
and for the shear
\begin{equation}
\sigma = - \frac{g}{\sqrt{3}} \ .
\end{equation}                                                                                                                           
These results exhibit some interesting features which are similar to those found for V1 (cp. section 3.1)
\begin{description}
\item{$\bullet$} The large-scale (relativistic) corrections are determined by only one function $g(t, r)$ which is proportional to the strength $K_1$ of the energy flow, vanishes, in accordance with the NRCM, at the coordinate origin and becomes singular for $r \to K^{-1/2}$ in case of a positive spatial curvature.
\item{$\bullet$} The radial pressure vanishes.
\item{$\bullet$} The transversal pressure turns out to be negative.
\item{$\bullet$} The spatial curvature shows no relativistic correction.
\item{$\bullet$} The electric part $E_{\mu\nu}$ of the Weyl tensor, which may be written as
\begin{equation}
E_{\mu\nu} = E (s_\mu s_\nu - \frac{1}{3} h_{\mu\nu})
\end{equation}
is different from zero. 

From the shear evolution equation, which in our case reads (cp. eq. (44) in [39]; note the different normalization of   $\sigma$)
\begin{equation}
\sqrt{3} \dot{\sigma} + \sigma^2 + \frac{2}{\sqrt{3}} \Theta \sigma = - (E + \frac{1}{2} p_t)  
\end{equation}
we obtain
\begin{equation}
E = - \frac{\dot{a}}{a} g - \frac{3}{2} g^2 \ .
\end{equation}
\end{description}

What about the sign of the energy density $\rho$?  According to (64) and (68) $\rho$ is always negative in the accelerating phase of the Universe.  In the decelerating phase and for $K > 0$ $\rho$ is positive for small r but then it turns over to become negative for large r and going to $-\infty$ for $r \to K^{-1/2}$.  So the situation is very different from the one found for model V1.

For our solutions (64)--(70) we may check again all the evolution and constraint equations for the kinematical and matter variables following from the Einstein equations. Because of the large number of these equations for a shearing anisotropic fluid (see [40]) we have restricted these checks to the two conservation equations
\begin{description}
\item{$\bullet$} The local energy conservation equation (62),
\item{$\bullet$} the momentum conservation equation, which reads in our case (see eq. (35) in [38])
\begin{equation}
\dot{q} + 2 q \left( 2 \frac{\dot{a}}{a} - g \right) - \frac{2 p_t}{r B} = 0 \ .
\end{equation}
\end{description}
Both are identically satisfied.

\section{Global averaging, backreaction and the size of the large-scale corrections}

Because we have found for both of our models V1 and V2 inhomogeneous solutions of Einstein's equations we should perform a global averaging before we are able to compare our findings with observational results. For reasons of simplicity we consider in the following exclusively model V2.

\bigskip
\noindent
{\bf Spatial averaging} of one scalar field  $\psi$ over a compact domain ${\cal D}$ with volume $V_D$ is defined by
\begin{equation}
<\psi >_{\cal D} : = \frac{1}{V_{\cal D}} \int_{\cal D} \psi (t,r) J (t,r) d^3 x
\end{equation}
where $J$ denotes the square root of the 3-metric determinant. 
          
\subsection{Backreaction}

Due to the noncommutativity between time-derivative and spatial averaging the evolution equations for the averaged kinematical and matter variables are different from the corresponding local equations. The difference can be described by additional source terms, called backreaction terms (see [7]). Most interesting is the backreaction term appearing in the averaged Raychaudhuri-Ehlers equation
\begin{eqnarray}
< \theta >^\cdot &+&  \frac{1}{3} < \theta >^2 = - 2 <\sigma >^2 - \frac{1}{2} (<\rho > + 3 <p> ) \nonumber\\
&+& \frac{2}{3} < (\theta - < \theta > )^2 > - 2 < (\sigma  - <\sigma > )^2| \ .
\end{eqnarray}

The backreaction term (last line of (64)) describes a correction to the effective gravitational energy density  $\rho + 3p$. For the model V1 the backreaction takes the form  $2/3 < (f - < f >)^2$. It is therefore a negative (dark energy) correction to the effective gravitational energy density.  On the other hand, for model V2 the backreaction vanishes due to a cancellation between the fluctuations of expansion rate and shear.

\subsection{Size of the relativistic corrections}

In the present paper we will focus our attention on small-sized relativistic corrections. Hence we restrict our considerations to model V2 and 
to the case $K(t) < 0$, relevant for the present day Universe (see  [3]).  Then we consider for the averaging domain in (75) a sphere of radius $R$ and take the limit $R \to \infty$. For that we need the asymptotic behavior of the following integrals
\begin{equation}
\int^R_0 \frac{r^{2+n}}{(1 + |K| r^2)^{\frac{1+n}{2}}} \mathop{\sim}_{R\to \infty} \frac{1}{2} \frac{R^2}{|K|^{\frac{1+n}{2}}}\ .
\end{equation}

Hence we obtain for the most important effective quantities
\begin{description}
\item{$\bullet$}  The scale factor, defined by
\begin{equation}
a_\infty (z):= \lim_{R\to \infty} \left(\frac{V_{\cal D}(z)}{V_{\cal D} (0)}\right)^{1/3}
\end{equation}
takes the form
\begin{equation}
a_\infty (z) = \frac{1}{1+z} \left( \frac{k(0)}{k(z)} \right)^{1/6}
\end{equation}

\item{$\bullet$} Expansion rate
\begin{equation}
H_\infty : = \frac{1}{3} < \Theta >_\infty = H_0 \left( h(z) + \frac{1}{3} \frac{k_1}{k(z)} (1 + z)^3 \right) \ .
\end{equation}
By using (79) we obtain the identity
\begin{equation}
H_\infty = \frac{\dot{a}_\infty}{a_\infty}
\end{equation}
which would not hold for a finite averaging radius $R$.
\end{description}

To illustrate the size of the relativistic corrections, we consider the case of scenario 2 from [1], reproduced in subsection 2.1 of the present paper.  In this case the correction for the scale factor is negligible. We get for the highest considered z-value, $z = 1.965$
\begin{equation}
\left( \frac{k(0)}{k(1.965)}\right)^{1/6} = 1.0027
\end{equation}
and for the corresponding effective Hubble function
\begin{equation}
H_\infty (1.965) = 189.97 - 1.17 = 188.80 \ .
\end{equation}
                             
The first number is the NRCM result and the second number gives the relativistic correction. 

The results (82) and (83) confirm a general statement by R\"as\"anen [41] that redshift and average expansion rate remain close to their background values if the metric and its first derivatives are close to the FLRW case.                                                                                                                                                       

To be complete we note also the results for the averaged transversal pressure
\begin{equation}
< p_t >_\infty = - 3 H^2_0 \frac{k^2_1 (1+z)^6}{k^2(z)}
\end{equation}
and the magnitude $E$ of the electric Weyl tensor
\begin{equation}
< E >_\infty = H^2_0 h(z) \frac{k_1}{k(z)} (1 + z)^3 + \frac{1}{2} < p_t >_\infty \ .
\end{equation}
We recall that the curvature function $K(t)$ experiences no correction.

\bigskip
\noindent
{\bf We conclude:} The relativistic corrections for scenario 2 from [1] are small for the considered z-values 
$z < 2$.

\section{Concluding remarks}
In this paper we have constructed two different relativistic generalizations of a nonrelativistic cosmological model with dynamical curvature (NRCM) whose solutions are fixed by three constants (initial conditions). These relativistic models rely on two inequivalent extensions of the FLRW  metric  containing besides a time dependent curvature function $K(t)$  a scale factor $a(t)$ as free functions. The corresponding Einstein equations are supposed to contain an energy momentum tensor with nontrivial pressure terms and energy flow. Then we have required that the now inhomogeneous solutions of the EEs agree in leading order at small distances with those of the NRCM. In technical terms this has been achieved by the demand of vanishing isotropic pressure and its first derivative at $x: = r^2 = 0$. In conclusion $a(t)$ and $K(t)$ will agree with their counterparts in the NRCM. Hence they are fixed by the three constants appearing in the NRCM and, therefore, we have obtained exact analytic solutions of the EEs for each of the two relativistic models.

In subsection 2.1 we have shown that the nonrelativistic version of our model can describe satisfactorily the cosmic chronometer data for the expansion rate. The very small value chosen for the constant $k$   leads at least for the model V2 to negligible inhomogeneities, to negligible relativistic corrections and predicts an almost constant negative curvature function for redshifts  $z < 2$.

Our models are different from the LTB-models as well as from the Stephani models. Whereas in the former case the spatial curvature depends on the spatial coordinates only our models share with the Stephani models the property of having a time-dependent spatial curvature of either sign.  Stephani models are characterized by a perfect fluid with a homogeneous energy density, an inhomogeneous pressure and an accelerating fluid flow. In contrast our models rely on an imperfect fluid (nontrivial energy flow) with energy density and pressure being inhomogeneous and the fluid flow is geodesic. But the basic difference between our models and the Stephani models is the mechanism causing the accelerated expansion of the Universe.  For that Stephani models need a large inhomogeneity.  For our models it is sufficient to have an energy flow of small magnitude which drives the accelerated expansion and, besides, leads only to a small inhomogeneity in the case of model V2.

To proceed with the models presented in this paper one should consider the following open problems:
\begin{description}
\item{$\bullet$} Look for at least plausibility arguments that each of our models can be viewed as a local average of some more fundamental cosmological model.

\item{$\bullet$}	Perform the global averaging procedure for both models V1 and V2 and for both signs of the curvature function.

\item{$\bullet$} To decide whether our models can be more than toy models they should pass more cosmological tests. In particular one should elaborate the distance-redshift relation for the averaged dynamics for both models.

Besides expansion rate and curvature function treated in the present paper we have until now only found the stationary solution of the EEs (for the EMT (2)) in non-comoving spherical coordinates given by one nonlinear ordinary differential equation for the gravitational potential [42]. The corresponding weak field limit [20] describes successfully galactic halos as shown in [18].
\end{description}

\section*{Acknowledgements}

I'm grateful to Thomas Buchert and the anonymous referee for discussion, interesting questions and valuable hints.

\section*{References}

\begin{description}

\item{[1]} P. Stichel, Dynamical curvature in a nonstandard cosmological model, arXiv:1712.03124
\item{[2]} D. Huterer and D.L. Shafer, Dark energy two decades after: Observables, probes, consistency 
      tests, Rep. Prog. Phys. 81, 016901 (2018)
\item{[3]} T. Buchert et al., Observational Challenges for the standard FLRW Model,
       Int. J. Mod. Phys. D 25, 1630007 (2016)
\item{[4]} S. Capozziello et al., Model independent constraints on dark energy evolution from low-redshift  observations, arXiv: 1806.03943
\item{[5]} C. Filho and E. Barboza Jr., Constraints on kinematic parameters at $z \ne 0$, arXiv: 1704.08089
\item{[6]} T. Buchert et al., Is there proof that backreaction of inhomogeneities is irrelevant in cosmology ?
       Class. Quant. Grav. 32, 215021 (2015).
\item{[7]} T. Buchert, Dark Energy from structure: s status report, Gen. Rel. Grav. 40, 467 (2008).
\item{[8]} T. Buchert and M. Carfora,  On the curvature of the present-day Universe, Class. Quant. Grav. 25, 195001 (2008).
\item{[9]} J. Larena et al., Testing backreaction effects with observations, Phys. Rev. D 79, 083011 (2009)
\item{[10]} K. Bolejko, Relativstic numerical cosmology with Silent Universes, arXiv:1708.09143.
\item{[11]} H. Stephani, \"Uber L\"osungen der Einsteinschen Feldgleichungen, die sich in einen f\"unfdimensionalen flachen Raum einbetten lassen, Commun. math. Phys. 4, 137 (1967).
\item{[12]} A. Krasinski, Inhomogeneous Cosmological Models, University Press , Cambridge (1997).
\item{[13]} M. Dabrowski, Isometric embedding of the spherically symmetric Stephani universe: Some explicit examples, J. Math. Phys. 34, 1447 (1993)
\item{[14]} R. Barrett and C. Clarkson, Undermining the cosmological principle: almost isotropic observations in inhomogeneous cosmologies, Class.Quant. Grav. 17, 5047 (2000)
\item{[15]} A. Balcerzak et al., Critical assessment o f some inhomogeneous pressure Stephani models, 
Phys. Rev.. D 91, 083506 (2015).
\item{[16]} Yen Ching Ong et al., Stephani Cosmology: Entropically Viable But Observationally Challenged, Eur. Phys. J. C78, 405 (2018)
\item{[17]} P.C. Stichel and W.J. Zakrzewski, Can cosmic acceleration be caused by exotic massless particles, 
      Phys. Rev. D 80, 083513 (2009).
\item{[18]} P.C. Stichel and W.J. Zakrzewski, Nonstandard approach to Gravity for the Dark Sector of the 
     Universe,  Entropy 15, 559 (2013).
\item{[19]} P.C. Stichel, Cosmological model with dynamical curvature, arXiv:1601.07030.
\item{[20]} P.C. Stichel and W.J. Zakrzewski, General relativistic, nonstandard model for the dark sector of the Universe, Eur. Phys .J. C, 75:9 (2015).
\item{[21]} F. Montanari and S. R\"as\"anen, Backreaction and FRW consistency conditions, JCAP 11, 032 (2017)
\item{[22]} F. Montanari, private communication.
\item{[23]} Clarkson et al., (Mis-)Interpreting supernovae observations in a lumpy universe, MNRAS 426, 1121 (2012)
\item{[24]} M. Dabrowski, A REDSHIFT-MAGNITUDE RELATION FOR NONUNIFORM PRESSURE UNIVERSES, ApJ, 447, 43 (1995)
\item{[25]} A.A. Coley and D.J. McManus, On spacetimes admitting shear-free, geodesic time-like    congruences, Class. Quant. Grav. 11, 1261 (1994)
\item{[26]} E.N. Glass, Shear-free gravitational collapse, J. Math. Phys. 20, 1508 (1979)
\item{[27]} O. Bergmann, A COSMOLOGICAL SOLUTION OF THE EINSTEIN EQUATIONS WITH HEAT FLOW,
        Phys. Lett. 82 A, 383 (1981)
\item{[28]} B. Modak, Cosmological Solution with Energy Flux, J. Astrophys. Astr. 5, 317 (1984)
\item{[29]} C. Barcelo and M. Visser, Twilight for the energy conditions ?, Int. J. Mod. Phys. D 11, 1553 (2002)
\item{[30]} A. Benoit-Levy and  G. Chardin, Introducing the Dirac-Milne universe, A\& A, 537, id A78 (2012)
\item{[31]} G. Manfredi et al., Cosmological structure formation with negative mass, Phys. Rev. D 98, 023514 (2018)
\item{[32]} J. Stelmach and I. Jakacka, Non-homogeneity-driven universe acceleration, Class. Quant. Grav. 18, 2643 (2001)
\item{[33]} W. Godlowski et al., Can the Stephani model be an alternative to FRW accelerating models?  Class. Quant. Grav. 21, 3953 (2004)
\item{[34]} M. Dabrowski and M. Hendry, THE HUBBLE DIAGRAM OF TYPE Ia SUPERNOVAE IN NON-UNIFORM PRESSURE UNIVERSES, Ap. J. 498, 67 (1998)
\item{[35]} A. Krasinski, Space-Times with Spherically Symmetric Hypersurfaces, Gen. Rel. Grav. 13, 1021 (1981)
\item{[36]} A. Paranjape and T. P. Singh, Explicit Cosmological Coarse Graining via Spatial Averaging, 
Gen. Rel. Grav. 40, 139 (2008)
\item{[37]} S. R\"as\"anen, Comment on ''Nontrivial Geometries: Bounds on the Curvature of the Universe'',
      Astropart. Phys. 30:216 (2008)
\item{[38]} L. Herrera, PHYSICAL CAUSES OF ENERGY-DENSITY INHOMOGENIZATION AND STABILITY OF ENERGY DENSITY HOMOGENEITY IN RELATIVISTIC SELF-GRAVITATING FLUIDS, Int. J. Mod. Phys. D 20, 1689 (2011).
\item{[39]} L. Herrera et al., On the stability of the shear-free condition, Gen. Rel. Grav. 42, 1585 (2010).
\item{[40]} G. Ellis and Henk van Elst, Cosmological Models, Cargese Lectures 1998.
\item{[41]} S. R\"as\"anen, Light propagation and the average expansion rate in near-FRW universes, 
        Phys. Rev. D 85, 083528 (2012I)
\item{[42]} P. Stichel, Anisotropic geodesic fluid in non-comoving spherical coordinates, 
        Journ.  Mod. Phys. 9, 207 (2018)
\item{[43]} H. Macpherson and D. Price, Einstein's Universe: Cosmological structure formation in numerical relativity, arXiv:1807.01711
\end{description}

\end{document}